\pdfoutput=1
\RequirePackage{ifpdf}
\ifpdf 
\documentclass[pdftex]{sigma}
\else
\documentclass{sigma}
\fi

\usepackage{mathrsfs}
\newcommand{\bbZ}{\mathbb{Z}}
\newcommand{\bbC}{\mathbb{C}}
\newcommand{\de}{\delta}
\newcommand{\ga}{\gamma}
\newcommand{\ka}{\kappa}
\newcommand{\ou}{\overline{u}}
\newcommand{\wu}{\widetilde{u}}
\newcommand{\wou}{\widetilde{\ou}}
\newcommand{\ov}{\overline{v}}
\newcommand{\wv}{\widetilde{v}}
\newcommand{\wov}{\widetilde{\ov}}
\newcommand{\PDE}{P$\Delta$E}

\numberwithin{equation}{section}

\newtheorem{Theorem}{Theorem}[section]
\newtheorem*{Theorem*}{Theorem}

 { \theoremstyle{definition}
\newtheorem{Definition}[Theorem]{Definition}

\newtheorem{Remark}[Theorem]{Remark} }

\begin{document}
\allowdisplaybreaks

\newcommand{\arXivNumber}{2201.11264}

\renewcommand{\PaperNumber}{032}

\FirstPageHeading

\ShortArticleName{Properties of the Non-Autonomous Lattice Sine-Gordon Equation}

\ArticleName{Properties of the Non-Autonomous\\ Lattice Sine-Gordon Equation:\\ Consistency around a Broken Cube Property}

\Author{Nobutaka NAKAZONO}

\AuthorNameForHeading{N.~Nakazono}

\Address{Institute of Engineering, Tokyo University of Agriculture and Technology,\\ 2-24-16 Nakacho Koganei, Tokyo 184-8588, Japan}
\Email{\href{mailto:nakazono@go.tuat.ac.jp}{nakazono@go.tuat.ac.jp}}
\URLaddress{\url{https://researchmap.jp/nakazono/}}

\ArticleDates{Received February 03, 2022, in final form April 14, 2022; Published online April 20, 2022}

\Abstract{The lattice sine-Gordon equation is an integrable partial difference equation on~${\mathbb Z}^2$, which approaches the sine-Gordon equation in a continuum limit. In this paper, we show that the non-autonomous lattice sine-Gordon equation has the consistency around a~broken cube property as well as its autonomous version. Moreover, we construct two new Lax pairs of the non-autonomous case by using the consistency property.}


\Classification{37K10; 39A14; 39A45}

\section{Introduction}\label{Introduction}
The sine-Gordon equation:
\begin{equation}\label{eqn:sg}
 \phi_{tt}-\phi_{xx}+\sin{\phi}=0,
\end{equation}
where $\phi=\phi(t,x)\in\bbC$ and $(t,x)\in\bbC^2$,
is well known as a motion equation of a row of pendulums hang from a rod and are coupled by torsion springs.
This equation is also known as a famous example of integrable systems.
In 1992, the following autonomous difference equation was found~\cite{bobenko1993discrete,volkov1992quantum}:
\begin{equation}\label{eqn:aut_lsg}
 \dfrac{u_{l+1,m+1}}{u_{l,m}}
 =\left(\dfrac{\ga-u_{l+1,m}}{1-\ga u_{l+1,m}}\right)\left(\dfrac{1-\ga u_{l,m+1}}{\ga-u_{l,m+1}}\right),
\end{equation}
where $u_{l,m}\in\bbC$, $(l,m)\in\bbZ^2$ and $\ga\in\bbC$,
which is a discrete analogue of equation~\eqref{eqn:sg} and therefore called lattice sine-Gordon (lsG) equation.
Note that it is not only equation~\eqref{eqn:aut_lsg} that is named lattice/discrete sine-Gordon equation.
Moreover, in 2018, the following non-autonomous version of the lsG equation was found \cite{kassotakis2018difference}:
\begin{equation}\label{eqn:nonaut_lsg}
 \dfrac{u_{l+1,m+1}}{u_{l,m}}
 =\left(\dfrac{p_{l+1}-q_{m}u_{l+1,m}}{q_{m}-p_{l+1}u_{l+1,m}}\right)
 \left(\dfrac{q_{m+1}-p_{l}u_{l,m+1}}{p_{l}-q_{m+1}u_{l,m+1}}\right),
\end{equation}
where $p_l,q_m\in\bbC$ are respectively arbitrary functions of $l$ and $m$.

A Lax pair is known as one of the most famous and important objects in the theory of integrable systems, which implies the integrability of differential/difference equations.
A Lax pair of equation~\eqref{eqn:aut_lsg} is already known \cite{bobenko1993discrete,HJN2016:MR3587455,joshi2021threedimensional,volkov1992quantum}, but that of equation~\eqref{eqn:nonaut_lsg} has not yet been reported.

In this paper, we focus on equation~\eqref{eqn:nonaut_lsg} and give its Lax pairs by using the consistency around a broken cube (CABC) property.
(See Appendix \ref{appendix:def_CABC} for the definition of CABC property).
The motivation for the discovery of the Lax pairs of equation~\eqref{eqn:nonaut_lsg} is as follows:
\begin{itemize}\itemsep=0pt
\item the autonomous lsG equation \eqref{eqn:aut_lsg} has the CABC property \cite{joshi2021threedimensional};
\item by using the CABC property, a Lax pair of equation~\eqref{eqn:aut_lsg} was constructed in \cite{{joshi2021threedimensional}}.
\end{itemize}
From the facts above, we can expect that the non-autonomous lsG equation \eqref{eqn:nonaut_lsg} also has the CABC property and by using it a Lax pair of equation~\eqref{eqn:nonaut_lsg}, as well as equation~\eqref{eqn:aut_lsg}, can be constructed.
This prediction is correct, and these results are summarized in Section~\ref{subsection:main}.

\subsection{Main results}\label{subsection:main}
In this subsection, we present the main results of this paper.

Firstly, in Section~\ref{section:lattice_structure}, we shall give proofs of the following theorem.

\begin{Theorem}\label{theorem:CABC}
Equation~\eqref{eqn:nonaut_lsg} has the CABC property.
\end{Theorem}

See Appendix \ref{appendix:def_CABC} for the definition of CABC property.
Using the CABC property, we obtain the following theorems.

\begin{Theorem}\label{theorem:Lax_CABC_1}
The following system for the two-vector $\Phi_{l,m}${\rm :}
\begin{equation*}
 \Phi_{l+1,m}=L_{l,m}\Phi_{l,m},\qquad
 \Phi_{l,m+1}=M_{l,m}\Phi_{l,m},
\end{equation*}
with
\begin{gather*}
 L_{l,m}
 =\begin{pmatrix}p_lp_{l+1}\left(\dfrac{p_l-q_mu_{l,m}}{q_m-p_lu_{l,m}}\right)u_{l+1,m}-{p_l}^2&\kappa+{p_l}^2\vspace{1mm}\\ p_1p_{l+1}\left(\dfrac{p_l-q_mu_{l,m}}{q_m-p_lu_{l,m}}\right)u_{l+1,m}&0\end{pmatrix},\\
 M_{l,m}
 =\begin{pmatrix}\dfrac{{p_l}^2-{q_m}^2}{q_m-p_lu_{l,m}}&\dfrac{\kappa+{p_l}^2}{p_lu_{l,m}}\vspace{1mm}\\ p_1\left(\dfrac{p_l-q_mu_{l,m}}{q_m-p_lu_{l,m}}\right)&\dfrac{p_l}{u_{l,m}}-q_m\end{pmatrix},
\end{gather*}
where $\ka\in\bbC$ is a spectral parameter,
is a Lax pair of equation~\eqref{eqn:nonaut_lsg}, that is, the compatibility condition
\begin{equation*}
 L_{l,m+1}M_{l,m}=M_{l+1,m}L_{l,m}
\end{equation*}
gives equation~\eqref{eqn:nonaut_lsg}.
\end{Theorem}

The proof of Theorem \ref{theorem:Lax_CABC_1} is given in Section~\ref{subsection:lattice_structure_1}.

\begin{Theorem}\label{theorem:Lax_CABC_2}
The following system for the two-vector $\Psi_{l,m}${\rm:}
\begin{equation*}
 \Psi_{l+1,m}={\mathcal L}_{l,m}\Psi_{l,m},\qquad
 \Psi_{l,m+1}={\mathcal M}_{l,m}\Psi_{l,m},
\end{equation*}
with
\begin{equation*}
 {\mathcal L}_{l,m}
 =\begin{pmatrix}0&1\\\ka\left(\dfrac{p_l-q_mu_{l,m}}{q_m-p_lu_{l,m}}\right)u_{l+1,m}&0\end{pmatrix},\qquad
 {\mathcal M}_{l,m}
 =\begin{pmatrix}0&\dfrac{1}{u_{l,m}}\\\ka\left(\dfrac{p_l-q_mu_{l,m}}{q_m-p_lu_{l,m}}\right)&0\end{pmatrix},
\end{equation*}
where $\ka\in\bbC$ is a spectral parameter,
is a Lax pair of equation~\eqref{eqn:nonaut_lsg}, that is, the compatibility condition
\begin{equation*}
 {\mathcal L}_{l,m+1}{\mathcal M}_{l,m}={\mathcal M}_{l+1,m}{\mathcal L}_{l,m}
\end{equation*}
gives equation~\eqref{eqn:nonaut_lsg}.
\end{Theorem}

The proof of Theorem \ref{theorem:Lax_CABC_2} is given in Section~\ref{subsection:lattice_structure_2}.

\subsection{Notation and terminology}
For conciseness in the remainder of the paper, we adopt the following notation for an arbitrary function $x_{l,m}$:
\begin{equation}\label{eqn:bar_tilde terminology}
 x=x_{l,m},\qquad
 \overline{x}=x_{l+1,m},\qquad
 \widetilde{x}=x_{l,m+1},\qquad
 \widetilde{\overline{x}}=x_{l+1,m+1},
\end{equation}
and extend the notation to other iterates of $x$ as needed.

We write each lattice equation as the vanishing condition of a polynomial of four variables.
For example, the lsG equation \eqref{eqn:nonaut_lsg} is given by $Q\big(u,\overline{u},\widetilde{u},\widetilde{\overline{u}}\big)=0$, where
\begin{equation*}
 Q\big(u,\ou,\wu,\wou\big)
 =\wou(q_m-p_{l+1}\ou)\big(p_l-q_{m+1}\wu\big)-u(p_{l+1}-q_m\ou)\big(q_{m+1}-p_l\wu\big).
\end{equation*}
(Where convenient, we also use lattice equations in their equivalent rational forms.)
Note that, for conciseness, we omit the dependence of the polynomial~$Q$ on parameters.
We assume that any parameters in the polynomial take generic values and that the corresponding polynomial is irreducible.

Because of the association with a quadrilateral of~$\bbZ^2$, a lattice equation relating four vertex values is called a {\it quad-equation}.
By a small abuse of terminology, we will also refer to the corresponding function, whose vanishing condition gives the lattice equation, as a quad-equation.
Moreover, if the polynomial defining a quad-equation is quadratic in each variable, we especially refer to it as a {\it multi-quadratic quad-equation}.

\subsection{Outline of the paper}
This paper is organized as follows.
In Section~\ref{section:lattice_structure}, showing the lattice structures of equation~\eqref{eqn:nonaut_lsg} in two ways, we give the proofs of Theorems~\ref{theorem:CABC}--\ref{theorem:Lax_CABC_2}.
Some concluding remarks are given in Section~\ref{ConcludingRemarks}.
Moreover, in Appendix \ref{appendix:def_CABC}, we give the definition of consistency around a broken cube property.

\section{Lattice structures of the lsG equation (\ref{eqn:nonaut_lsg})}\label{section:lattice_structure}
In this section, showing two types of lattice structures of equation~\eqref{eqn:nonaut_lsg} we give the proofs of Theorems~\ref{theorem:CABC}--\ref{theorem:Lax_CABC_2}.
See Appendix~\ref{appendix:def_CABC} for the definition of CABC and tetrahedron properties.

\subsection{CABC property of equation~(\ref{eqn:nonaut_lsg}): I}\label{subsection:lattice_structure_1}
We here start by defining the system of {\PDE}s:
\begin{subequations}\label{eqns:CABC1}
\begin{align}
 &{\mathsf A}\bigl(u,\ou,\wu,\wou\bigl)
 =\dfrac{~\wou~}{u}-\left(\dfrac{p_{l+1}-q_{m}\ou}{q_{m}-p_{l+1}\ou}\right)
 \left(\dfrac{q_{m+1}-p_{l}\wu}{p_{l}-q_{m+1}\wu}\right)=0,
 \label{eqn:CABC1_A}\\
 &{\mathsf S}\big(u,\ou,\wv,\wov\big)
 =\dfrac{1}{~\wv~}-\dfrac{{p_l}^2}{\ka+{p_l}^2}+\dfrac{p_lp_{l+1}u(p_{l+1}-q_m \ou)(1-\wov)}{(\ka+{p_l}^2)(q_m-p_{l+1}\ou)}=0,
 \label{eqn:CABC1_S}\\
 &{\mathsf B}\big(u,v,\wv\big)
 =\wv-\dfrac{(\ka+{p_l}^2)(q_m-p_lu)+p_l({p_l}^2-{q_m}^2)uv}{p_l(p_l-q_mu)(q_m-p_lu+p_luv)}=0,
 \label{eqn:CABC1_B}\\
 &{\mathsf C}\big(u,\ou,v,\ov\big)
 =\ov-1-\dfrac{(q_m-p_lu)(\ka+{p_l}^2-{p_l}^2v)}{p_lp_{l+1}(p_l-q_mu)\ou v}=0,
 \label{eqn:CABC1_C}
\end{align}
\end{subequations}
where we have used the terminology given in equation~\eqref{eqn:bar_tilde terminology} for $u_{l,m}$ and $v_{l,m}$.
Note that equation~\eqref{eqn:CABC1_A} is exactly equivalent to equation~\eqref{eqn:nonaut_lsg}.

It is straightforward to confirm that the system \eqref{eqns:CABC1} has the CABC and tetrahedron properties.
The tetrahedron equations ${\mathsf K}_1={\mathsf K}_1\big(u,\ou,v,\wov\big)$ and ${\mathsf K}_2={\mathsf K}_2\big(u,\ou,\ov,\wv\big)$ are given by
\begin{gather*}
 {\mathsf K}_1
 =\dfrac{p_{l+1}\big(\wov-1\big)(p_{l+1}-q_m\ou)}{q_m-p_{l+1}\ou}-\dfrac{p_l\big(\ka+{q_m}^2\big)(p_l-q_mu)v}{\big(\ka+{p_l}^2\big) (q_m-p_lu)+p_l\big({p_l}^2-{q_m}^2\big)uv}+q_m
 =0,\\
 {\mathsf K}_2
 =\dfrac{\ka+{p_l}^2}{p_l\wv}-\dfrac{\big(\ka+{q_m}^2\big)u}{q_m-p_{l+1}\ou+p_{l+1}\ou\,\ov}-p_l+q_mu
 =0.
\end{gather*}
Therefore, Theorem~\ref{theorem:CABC} holds.

Moreover, from the system \eqref{eqns:CABC1} we obtain the following equation given only by the variable~$v_{l,m}$:
\begin{gather}
 \big({p_{l+1}}^2(1-\ov)\big(1-\wov\big)v-{p_l}^2(1-v)(1-\wv)\wov\big)
 \big({p_l}^2(1-v)(1-\wv)\ov-{p_{l+1}}^2(1-\ov)\big(1-\wov\big)\wv\big)\notag\\
 \quad
 {}-{q_m}^2\big(\ka+{p_l}^2+{p_{l+1}}^2\big)\big(v\ov-\wv\,\wov\big)^2
 -{q_m}^2\big({p_l}^2-{p_{l+1}}^2\big)\big(\ov\,\wv-v\wov\big)\big(v\ov-\wv\,\wov\big)\notag\\
 \quad
 {}+\ka^2\big(v-\wov\big)(\ov-\wv)
 +\ka {p_l}^2(1-v)(1-\wv)\big(v\ov-2\ov\,\wov+\wv\,\wov\big)\notag\\
 \quad
 {}+\ka {p_{l+1}}^2(1-\ov)\big(1-\wov\big)\big(v\ov-2v\wv+\wv\,\wov\big)
 +\ka {q_m}^2\big(v+\ov-\wv-\wov\big)\big(v\ov-\wv\,\wov\big)\notag\\
 \quad{} +{q_m}^2{p_l}^2(1+v\wv)\big(\ov-\wov\big)\big(v\ov-\wv\,\wov\big)
 +{q_m}^2{p_{l+1}}^2\big(1+\ov\,\wov\big)(v-\wv)\big(v\ov-\wv\,\wov\big)
 =0,\label{eqn:CABC1_A'}
\end{gather}
which is assigned on the top face of each broken cube (see Figure~\ref{fig:cubeABCS}).
See the proof of Theorem~2 in \cite{joshi2021threedimensional} for details on how to derive a difference equation given only by the variable $v_{l,m}$ from a~system of {\PDE}s which has the CABC property.
Note that the system of equations \eqref{eqn:CABC1_S}--\eqref{eqn:CABC1_C} can also be regarded as a B\"acklund transformation from equation~\eqref{eqn:CABC1_A} to equation~\eqref{eqn:CABC1_A'}.

\begin{Remark}
The system (39) in \cite{joshi2021threedimensional}, which implies the CABC property of equation~\eqref{eqn:aut_lsg}, can be obtained from the system \eqref{eqns:CABC1} with the following specialization and transformation:
\begin{gather*}
 p_l=\ga,\qquad q_m=1,\qquad
 (u_{l,m},v_{l,m})\mapsto \big(u_{l,m},\ga^{-1}v_{l,m}\big).
\end{gather*}
\end{Remark}

Next, we construct the Lax pair in Theorem~\ref{theorem:Lax_CABC_1} through a method that parallels the well-known method for constructing a Lax pair using the consistency around a cube (CAC) property \cite{BS2002:MR1890049,NijhoffFW2002:MR1912127,WalkerAJ:thesis}.
Substituting
\begin{equation*}
 v_{l,m}=\dfrac{F_{l,m}}{G_{l,m}},
\end{equation*}
into the equations \eqref{eqn:CABC1_B} and \eqref{eqn:CABC1_C}
and separating the numerators and denominators of the resulting equations,
we obtain the following linear systems:
\begin{subequations}\label{eqns:CABC1_FG}
\begin{align}
 &F_{l+1,m}
 =\de^{(1)}_{l,m}\left(\left(p_lp_{l+1}\left(\dfrac{p_l-q_mu}{q_m-p_lu}\right)\ou-{p_l}^2\right)F_{l,m}+\big(\kappa+{p_l}^2\big)G_{l,m}\right),\\
 &G_{l+1,m}=\de^{(1)}_{l,m}p_1p_{l+1}\left(\dfrac{p_l-q_mu}{q_m-p_lu}\right)\ou F_{l,m},\\
 &F_{l,m+1}=\de^{(2)}_{l,m}\left(\dfrac{{p_l}^2-{q_m}^2}{q_m-p_lu}F_{l,m}+\dfrac{\kappa+{p_l}^2}{p_lu}G_{l,m}\right),\\
 &G_{l,m+1}=\de^{(2)}_{l,m}\left(p_1\left(\dfrac{p_l-q_mu}{q_m-p_lu}\right)F_{l,m}+\left(\dfrac{p_l}{u}-q_m\right)G_{l,m}\right),
\end{align}
\end{subequations}
where $\de^{(1)}_{l,m}$ and $\de^{(2)}_{l,m}$ are arbitrary decoupling factors.
Then, letting
\begin{equation*}
 \Phi_{l,m}=\begin{pmatrix}F_{l,m}\\G_{l,m}\end{pmatrix},
\end{equation*}
and taking
\begin{equation*}
 \de^{(1)}_{l,m}=1,\qquad
 \de^{(2)}_{l,m}=1,
\end{equation*}
from the equations \eqref{eqns:CABC1_FG} we obtain the Lax pair in Theorem \ref{theorem:Lax_CABC_1}.

\subsection{CABC property of equation~(\ref{eqn:nonaut_lsg}): II}\label{subsection:lattice_structure_2}
In this subsection, we show another system of {\PDE}s which also gives the CABC property of equation~\eqref{eqn:nonaut_lsg}.
The process for demonstrating the result is exactly the same as that in Section~\ref{subsection:lattice_structure_1} and so, for conciseness, we omit detailed arguments.

The system of {\PDE}s
\begin{subequations}\label{eqns:CABC2}
\begin{align}
 &{\mathsf A}\bigl(u,\ou,\wu,\wou\bigl)
 =\dfrac{~\wou~}{u}-\left(\dfrac{p_{l+1}-q_{m}\ou}{q_{m}-p_{l+1}\ou}\right)
 \left(\dfrac{q_{m+1}-p_{l}\wu}{p_{l}-q_{m+1}\wu}\right)=0,\label{eqn:CABC2_A}\\
 &{\mathsf S}\big(u,\ou,\wv,\wov\big)
 =\dfrac{1}{~\wv~}-\dfrac{\ka(p_{l+1}-q_m\ou)u\wov}{q_m-p_{l+1}\ou}=0,\\
 &{\mathsf B}\big(u,v,\wv\big)=\wv-\dfrac{q_m-p_lu}{\ka(p_l-q_mu)uv}=0,\\
 &{\mathsf C}\big(u,\ou,v,\ov\big)=\ov-\dfrac{q_m-p_lu}{\ka(p_l-q_mu)\ou v}=0,
\end{align}
\end{subequations}
where equation~\eqref{eqn:CABC2_A} is exactly equivalent to equation~\eqref{eqn:nonaut_lsg},
has the CABC and tetrahedron properties.
The tetrahedron equations are given by
\begin{gather*}
 {\mathsf K}_1
 =\dfrac{~\wov~}{v}-\dfrac{(q_m-p_{l+1}\ou)(p_l-q_mu)}{(p_{l+1}-q_m\ou)(q_m-p_lu)}=0,\\
{\mathsf K}_2=\dfrac{\ou}{u}-\dfrac{\wv}{\ov}=0,
\end{gather*}
and the equation represented only by the variable $v_{l,m}$ is given by
\begin{gather*}
 v\ov+\wv\wov
 -\dfrac{p_l(1+\ka v\wv)\ov\,\wov}{p_{l+1}\big(1+\ka\ov\,\wov\big)}
 -\dfrac{p_{l+1}\big(1+\ka\ov\,\wov\big)v\wv}{p_l(1+\ka v\wv)}
 -\dfrac{\ka {q_m}^2\big(v\ov-\wv\,\wov\big)^2}{p_lp_{l+1}(1+\ka v\wv)\big(1+\ka\ov\,\wov\big)}=0.
\end{gather*}
Therefore, the system \eqref{eqns:CABC2} gives another proof of Theorem \ref{theorem:CABC}.

Moreover, the Lax pair in Theorem \ref{theorem:Lax_CABC_2} can be obtained by using the system \eqref{eqns:CABC2} in the same way as in Section~\ref{subsection:lattice_structure_1}.

\section{Concluding remarks}\label{ConcludingRemarks}
 In this paper, we have shown the CABC property of the non-autonomous lattice sine-Gordon equation \eqref{eqn:nonaut_lsg}.
Moreover, using the CABC property we have constructed two Lax pairs for equation~\eqref{eqn:nonaut_lsg}.

Equation \eqref{eqn:nonaut_lsg} has properties similar to those of the Hirota's dKdV equation \cite{CNP1991:MR1111648,kajiwara2008bilinearization,TGR2001:REMBLAY2001319}:
\begin{equation}\label{eqn:deauto_dkdv_u}
 u_{l+1,m+1}-u_{l,m}=\dfrac{q_{m+1}-p_l}{u_{l,m+1}}-\dfrac{q_m-p_{l+1}}{u_{l+1,m}},
\end{equation}
according to our recent series of studies.
The common properties are, for example, that they have the CABC property shown in \cite{joshi2021threedimensional} and in this paper, and that they are not included in the list of equations, which have the CAC property, in \cite{ABS2003:MR1962121,ABS2009:MR2503862,BollR2011:MR2846098,BollR2012:MR3010833}.
Also, in a recent paper by the author \cite{nakazono2021discrete}, it was found that equation~\eqref{eqn:deauto_dkdv_u} has a special solution called the discrete Painlev\'e transcendent solution.
In fact, equation~\eqref{eqn:nonaut_lsg} also has a special solution of the same type.
This result will be reported in a forthcoming publication.
It is expected that there are many more equations besides the equations \eqref{eqn:nonaut_lsg} and \eqref{eqn:deauto_dkdv_u} that have these common properties.
We plan to derive equations with such properties in a future project.

\appendix

\section{Consistency around a broken cube property}\label{appendix:def_CABC}
In this appendix, we recall the definition of consistency around a broken cube (CABC) property.
We refer the reader to \cite{joshi2021threedimensional} for detailed information about this property.

\begin{figure}[hbtp]\centering
 \includegraphics[width=0.4\textwidth]{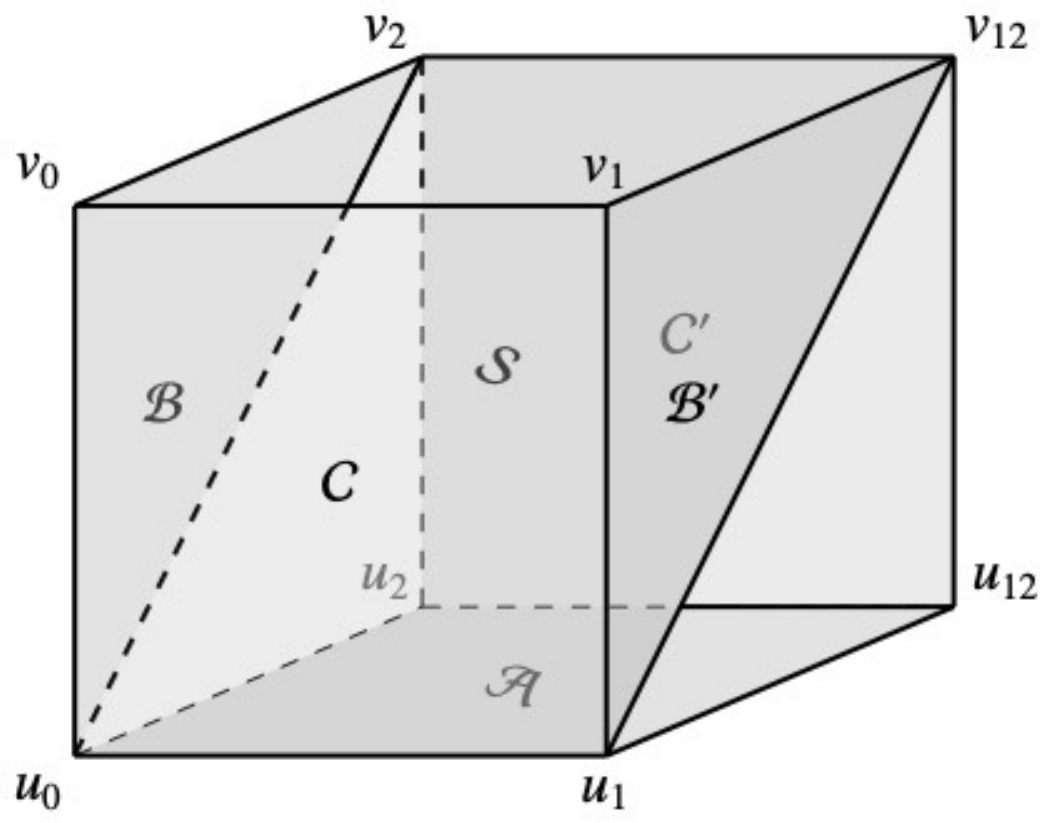}
\caption{A cube with three quadrilateral faces labelled by ${\mathcal A}$, ${\mathcal C}$ and ${\mathcal C}'$, an interior diagonal quadrilateral labelled by ${\mathcal S}$ and triangular domains labelled as ${\mathcal B}$ and ${\mathcal B}'$. Note that primes denote domains on parallel faces.}\label{fig:cubeABCS}
\end{figure}

We assign the following eight variables:
\begin{equation*}
 u_0,u_1,u_2,u_{12},v_0,v_1,v_2,v_{12}\in\bbC,
\end{equation*}
to vertexes of the cube as shown in Figure~\ref{fig:cubeABCS}.
In contrast to the usual procedure assumed for proving the consistency around a cube (CAC) property \cite{ABS2003:MR1962121,ABS2009:MR2503862,BollR2011:MR2846098,BollR2012:MR3010833,NCWQ1984:MR763123,NQC1983:MR719638,NW2001:MR1869690,nimmo1998integrable,QNCL1984:MR761644}, we do not assign a quad-equation to each face of the cube.
Instead, we describe a system of equations on the cube, which may (i) vary with each face; (ii) become a triangular equation, i.e., those relating only three vertex values, on certain faces; and, (iii) involves vertices of a quadrilateral given by an interior diagonal slice of the cube.

Three of the quad-equations occur on the bottom, front and back faces of the cube, while the fourth one occurs in the interior of the cube as a diagonal slice.
Each triangular domain occurs as a half of the left or right face of the cube.
See Figure \ref{fig:cubeABCS}.
We will refer to this configuration as a \textit{broken cube}.

Correspondingly, we define polynomials of 4 variables $\mathcal A, \mathcal S, \mathcal C, \mathcal C'\colon \bbC^4\to \bbC$ and those of 3 variables $\mathcal B, \mathcal B'\colon \bbC^3\to \bbC$, such that $\mathcal B$ and $\mathcal B'$ written as functions of $(x, y, z)$ satisfy
\begin{enumerate}\itemsep=0pt
\item[1)] $\deg_x{\mathcal B}\geq1$, $\deg_y{\mathcal B}=\deg_z{\mathcal B}=1$;
\item[2)] the equation ${\mathcal B}=0$ can be solved for $y$ and $z$, and each solution is a rational function of the other two arguments.
\end{enumerate}
With the labelling of vertices given in Figure \ref{fig:cubeABCS}, we denote the system of six corresponding equations by
\begin{subequations}\label{eq:face-eqns_localABCS}
\begin{alignat}{3}
 &{\mathcal A}(u_0,u_1,u_2,u_{12})=0,\qquad &&{\mathcal S}(u_0,u_1,v_2,v_{12})=0,&\\
 &{\mathcal B}(u_0,v_0,v_2)=0,\qquad &&{\mathcal B}'(u_1,v_1,v_{12})=0,& \\
 &{\mathcal C}(u_0,u_1,v_0,v_1)=0, \qquad && {\mathcal C}'(u_2,u_{12},v_2,v_{12})=0.&
\end{alignat}
\end{subequations}
The following definition describes how consistency holds for this system of equations.

\begin{Definition}[CABC property]\label{def:CABC_local}\rm
Let $\{u_0,u_1,u_2,v_0\}$ be given initial values.
Using equations~\eqref{eq:face-eqns_localABCS},
we can express the variable $v_{12}$ as a rational function in terms of the initial values in 3 ways.
When the 3 results for $v_{12}$ are equal, the system of equations \eqref{eq:face-eqns_localABCS} is said to be \emph{consistent around a broken cube} or to have the \emph{consistency around a broken cube} (CABC) property.
In this case, we refer to the configuration of quadrilaterals and triangular domains associated with the polynomials $\mathcal A$, $\mathcal S$, $\mathcal C$, $\mathcal C'$, $\mathcal B$, $\mathcal B'$ as a {\it CABC cube}.
\end{Definition}

Other equations arise from interrelationships between the above equations on the broken cube.
For example, an equation arises on the top face, parallel to $\mathcal A$.
It is also useful to note equations that relate three vertices on a face to a vertex on the opposite face.
The following definition of such equations uses terminology analogous to existing ones in the literature on the CAC property.

\begin{Definition}[tetrahedron property]\label{def:tetrahedron_local}\rm
A CABC cube is said to have a \emph{tetrahedron property},
if there exist quad-equations ${\mathcal K}_1$ and ${\mathcal K}_2$ satisfying
\begin{equation*}
 {\mathcal K}_1(u_0,u_1,v_0,v_{12})=0,\qquad
 {\mathcal K}_2(u_0,u_1,v_1,v_2)=0.
\end{equation*}
In this case, each of the equations ${\mathcal K}_1=0$ and ${\mathcal K}_2=0$ is referred to as a \emph{tetrahedron equation}.
\end{Definition}

By interpreting each vertex value as an iterate of a function in an appropriate way, we can interpret the above equations as \PDE s.
In particular, we use the terminology given in equation~\eqref{eqn:bar_tilde terminology} for $u_{l,m}$ and $v_{l,m}$ to give the following definition of \PDE s.

\begin{Definition}[CABC and tetrahedron properties for a system of {\PDE}s]
Define the \PDE s
\begin{equation}\label{eqn:general_P1234}
 {\mathsf A}\bigl(u,\ou,\wu,\wou\bigl)=0,\qquad
 {\mathsf S}\big(u,\ou,\wv,\wov\big)=0,\qquad
 {\mathsf B}\big(u,v,\wv\big)=0,\qquad
 {\mathsf C}\big(u,\ou,v,\ov\big)=0,
\end{equation}
which give the following equations around each elementary cubic cell in~$\bbZ^3$:
\begin{subequations}\label{eqn:general_ASBBCC}
\begin{alignat}{3}
 &\mathcal A={\mathsf A}\bigl(u,\ou,\wu,\wou\bigl)=0,
\qquad
 &&\mathcal S={\mathsf S}\big(u,\ou,\wv,\wov\big)=0,&\\
 &{\mathcal B}={\mathsf B}\big(u,v,\wv\big)=0,
\qquad
 &&{\mathcal B}'={\mathsf B}\big(\ou,\ov,\wov\big)=0,& \\
 &{\mathcal C}={\mathsf C}\big(u,\ou,v,\ov\big)=0,\qquad
 &&{\mathcal C}'={\mathsf C}\big(\wu,\wou,\wv,\wov\big)=0.&
\end{alignat}
\end{subequations}
Then, the system \eqref{eqn:general_P1234} is said to have the CABC property if Definition~\ref{def:CABC_local} holds for the equations~\eqref{eqn:general_ASBBCC}.
We also transfer the definition of tetrahedron properties to {\PDE}s corresponding to $\mathcal K_j$, $j=1,2$, in the obvious way.
Moreover, the {\PDE}
\begin{equation*}
 {\mathsf A}\bigl(u,\ou,\wu,\wou\bigl)=0
\end{equation*}
will be described as having the CABC property, if the system \eqref{eqn:general_P1234} has the CABC property.
\end{Definition}

\begin{Remark}
Note that equations \eqref{eqn:general_P1234} are not necessarily autonomous. They may contain parameters that evolve with $(l, m)$.
\end{Remark}

\subsection*{Acknowledgment}
This research was supported by a JSPS KAKENHI Grant Number JP19K14559.

\pdfbookmark[1]{References}{ref}
\LastPageEnding


\begin{thebibliography}{99}
\footnotesize\itemsep=0pt

\bibitem{ABS2003:MR1962121}
Adler V.E., Bobenko A.I., Suris Yu.B., Classification of integrable equations on
 quad-graphs. {T}he consistency approach, \href{https://doi.org/10.1007/s00220-002-0762-8}{\textit{Comm. Math. Phys.}}
 \textbf{233} (2003), 513--543, \href{https://arxiv.org/abs/nlin.SI/0202024}{arXiv:nlin.SI/0202024}.

\bibitem{ABS2009:MR2503862}
Adler V.E., Bobenko A.I., Suris Yu.B., Discrete nonlinear hyperbolic equations:
 classification of integrable cases, \href{https://doi.org/10.1007/s10688-009-0002-5}{\textit{Funct. Anal. Appl.}} \textbf{43}
 (2009), 3--17, \href{https://arxiv.org/abs/0705.1663}{arXiv:0705.1663}.

\bibitem{bobenko1993discrete}
Bobenko A., Kutz N., Pinkall U., The discrete quantum pendulum, \href{https://doi.org/10.1016/0375-9601(93)90965-3}{\textit{Phys.
 Lett.~A}} \textbf{177} (1993), 399--404.

\bibitem{BS2002:MR1890049}
Bobenko A.I., Suris Yu.B., Integrable systems on quad-graphs, \href{https://doi.org/10.1155/S1073792802110075}{\textit{Int. Math.
 Res. Not.}} \textbf{2002} (2002), 573--611, \href{https://arxiv.org/abs/nlin.SI/0110004}{arXiv:nlin.SI/0110004}.

\bibitem{BollR2011:MR2846098}
Boll R., Classification of 3{D} consistent quad-equations, \href{https://doi.org/10.1142/S1402925111001647}{\textit{J.~Nonlinear
 Math. Phys.}} \textbf{18} (2011), 337--365, \href{https://arxiv.org/abs/1009.4007}{arXiv:1009.4007}.

\bibitem{BollR2012:MR3010833}
Boll R., Corrigendum: {C}lassification of 3{D} consistent quad-equations,
 \href{https://doi.org/10.1142/S1402925112920015}{\textit{J.~Nonlinear Math. Phys.}} \textbf{19} (2012), 1292001, 3~pages.

\bibitem{CNP1991:MR1111648}
Capel H.W., Nijhoff F.W., Papageorgiou V.G., Complete integrability of
 {L}agrangian mappings and lattices of {K}d{V} type, \href{https://doi.org/10.1016/0375-9601(91)91043-D}{\textit{Phys. Lett.~A}}
 \textbf{155} (1991), 377--387.

\bibitem{HJN2016:MR3587455}
Hietarinta J., Joshi N., Nijhoff F.W., Discrete systems and integrability,
 \textit{Cambridge Texts in Applied Ma\-the\-matics}, \href{https://doi.org/10.1017/CBO9781107337411}{Cambridge University Press},
 Cambridge, 2016.

\bibitem{joshi2021threedimensional}
Joshi N., Nakazono N., On the three-dimensional consistency of {H}irota's
 discrete {K}orteweg--de {V}ries equation, \href{https://doi.org/10.1111/sapm.12421}{\textit{Stud Appl. Math.}}
 \textbf{147} (2021), 1409--1424, \href{https://arxiv.org/abs/2102.00684}{arXiv:2102.00684}.

\bibitem{kajiwara2008bilinearization}
Kajiwara K., Ohta Y., Bilinearization and {C}asorati determinant solution to
 the non-autonomous discrete {K}d{V} equation, \href{https://doi.org/10.1143/JPSJ.77.054004}{\textit{J.~Phys. Soc. Japan}}
 \textbf{77} (2008), 054004, 9~pages, \href{https://arxiv.org/abs/0802.0757}{arXiv:0802.0757}.

\bibitem{kassotakis2018difference}
Kassotakis P., Nieszporski M., Difference systems in bond and face variables
 and non-potential versions of discrete integrable systems,
 \href{https://doi.org/10.1088/1751-8121/aad4c4}{\textit{J.~Phys.~A: Math. Theor.}} \textbf{51} (2018), 385203, 21~pages,
 \href{https://arxiv.org/abs/1710.11111}{arXiv:1710.11111}.

\bibitem{nakazono2021discrete}
Nakazono N., Discrete {P}ainlev\'e transcendent solutions to the multiplicative
 type discrete {K}d{V} equations, \textit{J.~Math. Phys.}, {t}o appear, \href{https://arxiv.org/abs/2104.11433}{arXiv:2104.11433}.

\bibitem{NijhoffFW2002:MR1912127}
Nijhoff F.W., Lax pair for the {A}dler (lattice {K}richever--{N}ovikov) system,
 \href{https://doi.org/10.1016/S0375-9601(02)00287-6}{\textit{Phys. Lett.~A}} \textbf{297} (2002), 49--58, \href{https://arxiv.org/abs/nlin.SI/0110027}{arXiv:nlin.SI/0110027}.

\bibitem{NCWQ1984:MR763123}
Nijhoff F.W., Capel H.W., Wiersma G.L., Quispel G.R.W., B\"acklund
 transformations and three-dimensional lattice equations, \href{https://doi.org/10.1016/0375-9601(84)90994-0}{\textit{Phys.
 Lett.~A}} \textbf{105} (1984), 267--272.

\bibitem{NQC1983:MR719638}
Nijhoff F.W., Quispel G.R.W., Capel H.W., Direct linearization of nonlinear
 difference-difference equations, \href{https://doi.org/10.1016/0375-9601(83)90192-5}{\textit{Phys. Lett.~A}} \textbf{97} (1983),
 125--128.

\bibitem{NW2001:MR1869690}
Nijhoff F.W., Walker A.J., The discrete and continuous {P}ainlev\'e {VI}
 hierarchy and the {G}arnier systems, \href{https://doi.org/10.1017/S0017089501000106}{\textit{Glasg. Math.~J.}} \textbf{43A}
 (2001), 109--123, \href{https://arxiv.org/abs/nlin.SI/0001054}{arXiv:nlin.SI/0001054}.

\bibitem{nimmo1998integrable}
Nimmo J.J.C., Schief W.K., An integrable discretization of a
 {$(2+1)$}-dimensional sine-{G}ordon equation, \href{https://doi.org/10.1111/1467-9590.00079}{\textit{Stud. Appl. Math.}}
 \textbf{100} (1998), 295--309.

\bibitem{QNCL1984:MR761644}
Quispel G.R.W., Nijhoff F.W., Capel H.W., van~der Linden J., Linear integral
 equations and nonlinear difference-difference equations, \href{https://doi.org/10.1016/0378-4371(84)90059-1}{\textit{Phys.~A}}
 \textbf{125} (1984), 344--380.

\bibitem{TGR2001:REMBLAY2001319}
Tremblay S., Grammaticos B., Ramani A., Integrable lattice equations and their
 growth properties, \href{https://doi.org/10.1016/S0375-9601(00)00806-9}{\textit{Phys. Lett.~A}} \textbf{278} (2001), 319--324,
 \href{https://arxiv.org/abs/0709.3095}{arXiv:0709.3095}.

\bibitem{volkov1992quantum}
Volkov A.Yu., Faddeev L.D., Quantum inverse scattering method on a spacetime
 lattice, \href{https://doi.org/10.1007/BF01015552}{\textit{Theoret. and Math. Phys.}} \textbf{92} (1992), 837--842.

\bibitem{WalkerAJ:thesis}
Walker A., Similarity reductions and integrable lattice equations, Ph.D.~Thesis, {U}niversity of Leeds, 2001.

\end{thebibliography}
\end{document}